\documentclass[preprint,unsortedaddress,amsmath,amssymb,aps]{revtex4-1}
\pdfoutput=1
\usepackage{graphicx}

\usepackage{color}
\usepackage[draft]{todonotes}   

\newcommand{\comment}[1]

\usepackage{dcolumn}
\usepackage{bm}

\begin{document}

\title{A mean field approach to model levels of consciousness from EEG recordings}

\author{Marco Alberto Javarone}
\email{marcojavarone@gmail.com}
\affiliation{Department of Mathematics, University College London, London, UK}

\author{Olivia Gosseries}
\affiliation{Coma Science Group, GIGA Consciousness University and University Hospital of Liege, Liege, Belgium}

\author{Daniele Marinazzo}
\affiliation{University of Ghent, Ghent, Belgium}

\author{Quentin Noirhomme}
\affiliation{Faculty of Psychology and Neuroscience Maastricht University, Maastricht, Netherlands}

\author{Vincent Bonhomme}
\affiliation{GIGA - Consciousness, Anesthesia and Intensive Care Medicine Laboratory University and CHU University Hospital of Liege, Liege, Belgium \\ Department of Anesthesia and Intensive Care Medicine CHU University Hospital of Liege and CHR Citadelle, Liege, Belgium}

\author{Steven Laureys}
\thanks{these two authors contributed equally}
\affiliation{Coma Science Group, GIGA Consciousness University and University Hospital of Liege, Liege, Belgium}

\author{Srivas Chennu}
\thanks{these two authors contributed equally}
\affiliation{University of Kent, Medway, UK \\ University of Cambridge, Cambridge, UK}

\date{\today}

\begin{abstract}
We introduce a mean-field model for analysing the dynamics of human consciousness. In particular, inspired by the Giulio Tononi's Integrated Information Theory and by the Max Tegmark's representation of consciousness, we study order-disorder phase transitions on Curie-Weiss models generated by processing EEG signals. The latter have been recorded on healthy individuals undergoing deep sedation. Then, we implement a machine learning tool for classifying mental states using, as input, the critical temperatures computed in the Curie-Weiss models.
Results show that, by the proposed method, it is possible to discriminate between states of awareness and states of deep sedation. Besides, we identify a state space for representing the path between mental states, whose dimensions correspond to critical temperatures computed over different frequency bands of the EEG signal. Beyond possible theoretical implications in the study of human consciousness, resulting from our model, we deem relevant to emphasise that the proposed method could be exploited for clinical applications.
\end{abstract}

\maketitle
\section{Introduction}
Consciousness is one of the most complex and fascinating phenomena in the brain, attracting the interest of a variety of scholars, spanning from neuroscientists to mathematicians, and from physicists to philosophers~\cite{tononi00,blackmore01,tegmark01,zhou01,fingelkurts1,fingelkurts2,penrose01,friston02}. In addition, consciousness, as well as other complex systems as those we find in biology, social science, finance and artificial intelligence~\cite{melaniemitchel01,kaneko01,flack01,vespignani01,rickles01,sole01,friston01}, has strongly benefited from the introduction of cross-disciplinary approaches.
Despite a huge number of investigations, a lot of its aspects and mechanisms still require to be clarified.
Given these observations, we focus on the challenge of quantifying consciousness, putting attention on the transition between mental states. So, we introduce a method for generating Curie-Weiss models~\cite{barra01,wolski01} from EEG signals, and then we analyse its outcomes by a machine learning classifier. As discussed later, although the spirit of this work is mostly theoretical, we conceive a framework that could support clinicians in some relevant tasks, e.g. in calibrating the optimal amount of anaesthetic for patients, and in assessing the cognitive conditions of unresponsive individuals.
Nowadays, a number of devices allow studying the brain structure and its dynamics. Usually, the choice of a specific tool reflects both clinical needs and patient conditions. Here, we use recordings obtained by EEG analyses for two main reasons, i.e. its cheaper cost compared to other technologies and its non-invasive nature. At the same time, it is worth to report that the EEG is less advanced than other diagnostic tools, as fMRI, that generate images of higher quality (e.g. higher spatial resolution). Notwithstanding, stimulated by the above reasons, we aim to improve as much as possible the value of the information content of EEG signals, concerning the dynamics of human consciousness.
Before moving to the proposed method, we very briefly introduce two seminal works on this topic, the Integrated Information Theory (IIT hereinafter) developed by Giulio Tononi~\cite{tononi01} and the Max Tegmark's manuscript on the physical representation of consciousness as a state of matter~\cite{tegmark01}. Both works contain ideas an observations that inspired us during our investigation.
The IIT is based on the core concept that human consciousness results from integrated information, generated by an ensemble of interacting elements. So, information emerges from collective action, and its content is extremely much richer than that one can obtain just by a simple summation of the individual contributions, like those provided by the elements belonging to the same ensemble. More in general, the concept of collective effect pervades the field of complex systems, as effectively explained by Anderson in 'More is different'~\cite{anderson01}.
Under that light, Tegmark proposed a computational description of consciousness~\cite{tegmark01}, trying to address the IIT by the language of Physics, and developing both a classical and a quantum representation of this phenomenon.
From his work~\cite{tegmark01}, we take into account the 'classical' description of IIT, achieved via the Ising model, where a 'conscious' regime emerges only within a very restricted range of values. Such restricted range refers to the collective phenomena occurring when a spin system gets close to its critical temperature, and it is in full agreement with the Damasio's observations~\cite{damasio01} on the conditions required for the reaching of homeostasis (i.e. some physical parameters, in the brain, have to be kept within a narrow range of values).
Therefore, following ideas and insights of the above-mentioned authors, we propose a method for building a mean-field model from EEG data. Then, we analyse its behaviour by implementing Monte Carlo simulations, whose outcomes are expected to provide the information we need to quantify the transitions between mental states of individuals undergoing deep sedation. 
In particular, we consider the critical temperature computed in the various realisations of the model, e.g. those achieved on varying the frequency of the EEG signal.
Details about the proposed model and the method for classifying mental states are provided in the next section. Here we take the opportunity for highlighting that, despite the increasing interest for modelling brain dynamics by networked approaches (e.g.~\cite{chialvo01,bassett01,bassett02,bassett03,bullmore01,dedomenico01,caldarelli01,diessen01,fallani01,srivas01,srivas02}), the present investigation is based on the modelling and the analysis of the distribution of electrical activity recorded in the scalp.
Moreover, to each frequency band of the signal, as $\delta$ for $1-4$ Hz and $\theta$ for $4-7$ Hz, corresponds a Curie-Weiss model whose interactions depend on the recorded phase differences across scalp locations.
It is worth to note that previous studies showed that the $\alpha$ ($8-12$ Hz) band can be useful for quantifying transitions between mental states (e.g.~\cite{srivas01}), as well as other signal components. At the same time, most of these works consider networks generated by avoiding to include 'weak' interactions. Notably, to remove weak interactions a threshold needs to be defined, and such practice has received some fair criticisms~\cite{papo01}. Remarkably, to the best of our knowledge, the definition of a suitable threshold is currently based only on rules of thumb.
So, it is important to remark that the proposed model does not require to filter out, or to cut off, weak interactions.
Eventually, let us observe that while from a neuroscience perspective the human consciousness might be investigated considering the full set of frequency bands of an EEG signal, those of major interest seem to be the $\delta$, the $\alpha$, and the $\beta$ band.
However, due to the influence that has been reported between the propofol, i.e. the drug administered in our individuals during the examination, and the behaviour of the $\beta$ band~\cite{boussen01,mccarthy01}, we decided to take into account only the bands $\delta$ and $\alpha$. A more detailed list of features, for analysing consciousness, can be found in~\cite{engemann01}.
Summarising, our goal is to quantify consciousness, looking also at potential clinical applications. Notably, the EEG signal, as currently processed, provides some information about the state of consciousness of a patient, but it has some limitations. For instance, assessing the level of unconsciousness, or understanding why some individuals report having been fully aware (even if, obviously, unable to communicate) during surgery, is currently difficult by inspecting only the EEG signal. Therefore, under the assumption that the latter might contain more information than those currently extracted, we propose a method to improve its content (see also~\cite{juel01}). Notably, in mathematical terms, the proposed model can be thought of as a more rich representation of the EEG signal, being mapped to a novel vector space that we define state space of mental states. In relation to that, the Curie-Weiss model represents the tool to generate that space of states, by identifying the critical temperatures of each individual during an examination. Here, while critical temperatures are computed to generate the state space of mental states, they have no meaning with what is occurring in the brain of individuals. It is also worth to remark that our choice of using a model (i.e. the Curie-Weiss) usually adopted to describe collective phenomena, as phase transitions, aims to build a direct link with the IIT framework, where the concept of collective behaviour is central.
The remainder of the paper is organised as follows: Section~\ref{sec:model} introduces the proposed model. Section~\ref{sec:results} shows results of numerical simulations. Eventually, Section~\ref{sec:conclusions} ends the paper providing an overall discussion on this investigation, from its goal to the main outcomes, and on some possible future developments.
\section{Model}\label{sec:model}
In this section, we describe a framework for classifying mental states, whose variation is represented as a phase transition.
In statistical mechanics, the most simple representation of phase transitions is achieved by the Ising model, and the latter has been used in~\cite{tegmark01} for showing how, in terms of information content, the set of states reachable by that model, at the critical temperature, contains suitable candidates for representing states of consciousness. 
Therefore, we aim to evaluate with data (i.e. EEG recordings) whether that theoretical insight can be exploited for quantifying consciousness and performing classification tasks. It is worth to add that the Ising model has been used also by other authors for investigating different dynamics of the brain (see for instance~\cite{chialvo01,bassett02,marinazzo02,bassett04,das01,abeyasinghe01}). 
As above mentioned, the EEG signal can be decomposed into frequency bands and different measures can be adopted for its analysis, usually selected according to specific needs. For our purposes, a particularly useful parameter is the weighted Phase Lag Index~\cite{srivas02} (wPLI hereinafter) that quantifies the correlation between pairs of sensors in the scalp. 
In general, considering two time series $a(t)$ and $b(t)$, the wPLI is defined as
\begin{equation}\label{eq:wpli}
wPLI = | \frac{\sum_{t=i}^n |im(P_{ab,t})| \text{ sgn } im(P_{ab,t})} {\sum_{t=i}^n | im(P_{ab,t})|} |
\end{equation}
\noindent where \textit{sgn} indicates the signum function, \textit{im} indicates the imaginary part, and $P_{ab,t}$ the complex cross-spectral density, of $a(t)$ and $b(t)$, at time $t$. Notably, given the power spectral densitities of the two signals $P_{aa}$ and $P_{bb}$, i.e. the distribution of the power across the frequency components of $a(t)$ and $b(t)$, respectively, the cross-spectral density quantifies the correlation between them.
Then, for each frequency band, the interactions of the resulting Curie-Weiss are computed by scalar multiplication of the wPLI index with the relative power of the signal.
A quick inspection of~\ref{eq:wpli} shows that the wPLI and the power of the signal are strongly correlated. However, we found beneficial to combine them for realising the mean-field model.
In doing so, inspired by the Tegmark's approach for studying the IIT by a simple physical system, and remaining in the land of the classical physics, we build a Curie-Weiss from EEG recordings assigning a spin to each sensor and computing interactions by the combination of the indexes above mentioned (i.e. wPLI and Power).
Thus, starting with randomly assigned values of spins ($\sigma \pm 1$), and quenching interactions~\cite{javarone01,barra01}, we study the dynamics of the system towards equilibrium.
Following this method, it is possible to compute a critical temperature for each frequency band of the EEG signal, as $T_c^{\alpha}$ for the $\alpha$ band.
However, since this signal varies over time, node interactions can vary as well. Here, actually, the variation of interactions is expected to be useful for detecting variations of mental states.
For instance, as reported in~\cite{srivas01}, networks built using the wPLI index show variations as individuals undergo sedation and then recover to their original conscious state. 
To tackle this aspect, the EEG signal is sampled into four different points labelled as $C$, $S$, $DS$, $R$, representing consciousness, sedation, deep sedation, and recovery, respectively. Mental states $C$ and $DS$ are both classified as states of consciousness, in agreement with ~\cite{tononi02}, while $S$ and $R$ are labelled as transition states. 
Also, $C$ and $DS$ can be viewed as two equilibrium states (although the deep sedation, i.e. $DS$, in our individuals has been induced by a drug).
Let us now proceed to the formal definition of spin interactions.
Recalling that the EEG signal is decomposed into $5$ main frequency bands and that we extract $4$ samples per recording, for each individual we can generate up to $20$ mean-field models.
Since the wPLI quantifies the correlation between pairs of nodes, we indicate with $wPLI^{x}_{i,j}$ the correlation between sensors $i$ and $j$ in the $x$th frequency band.
Accordingly, the interaction term $J$ reads
\begin{equation}\label{eq:link}
J^{x}_{i,j}(s) = P^{x}(s) \cdot wPLI^{x}_{i,j}(s) 
\end{equation}
\noindent with $s$ sample (or mental state) and $P^{x}(s)$ power of the $x$th band for that specific sample.
Thus, the Hamiltonian of the system is
\begin{equation}\label{eq:hamiltonian}
H = -\frac{1}{N} \sum_{i,j=1,j \neq i}^{N} J_{i,j}\sigma_i \sigma_j 
\end{equation}
\noindent with $N$ number of sensors and $\sigma$ spin assigned to them. 
While the spin is a quantum property of particles, it finds large utilisation in non-physical models, typically for representing binary features.
For instance, in social dynamics the spin can represent a binary opinion~\cite{galam01}, in evolutionary game theory a strategy~\cite{perc01,javarone02}, and in neural models it can indicate firing ($+1$) and resting ($-1$) states~\cite{amit01}.
The proposed model does not define explicitly a connection between the value of the spin and the underlying neural activity since the spin dynamics is implemented only to obtain theoretical insights about the activity distribution recorded in the scalp.
In particular, the studying of the order-disorder phase transitions, occurring in each model realisation, allows computing the set of critical temperatures $T_c$ associated to every mental state.
To this end, for each configuration (i.e. an individual in a given state), spin interactions $J$ are considered as \textit{quenched}, so one can analyse the dynamics of spins starting from a random distribution. 
Here, since our efforts are directed towards quantifying the human consciousness, we put the attention on the $\alpha$ and $\delta$ bands that, according to previous clinical studies (e.g.~\cite{tononi04,golkowski01}), seem to be quite relevant for investigating this complex phenomenon, as well as others as psychotic disorders~\cite{howells01}.
The described method allows observing the motion that individuals take in the space of mental states. Such motion is defined along a path on a bidimensional plane, whose axes are $(T_c^{\alpha},T_c^{\delta})$. Moreover, since recordings terminate once individuals recover their original cognitive state, the resulting path forms a closed cycle.
Then, we implement a Machine Learning tool to exploit the resulting paths (or cycles) for classification tasks, as for identifying the correct label of a point (e.g. $DS$) in the mental state space.
In particular, a classifier able to assign a label to each point takes as input vectors with components $(T_c^{\alpha},T_c^{\delta})$.
As a huge literature suggests, classifiers can be realised by many algorithms, e.g. neural networks~\cite{bengio01}. However, given the small size of our dataset, we implemented a Support Vector Machine (SVM hereinafter)~\cite{bengio01} ---see also~\ref{sec:appendix_a}.
Summarising, starting from EEG recordings, the proposed model generates bidimensional vectors, whose entries correspond to the critical temperatures computed in the $\alpha$ and $\delta$ frequency bands, state by state. These vectors constitute the input of an SVM designed to discriminate between the two states of consciousness $C$ and $DS$.
\section{Results}\label{sec:results}
\begin{figure}[h!]
\centering
\includegraphics[width=6.5in]{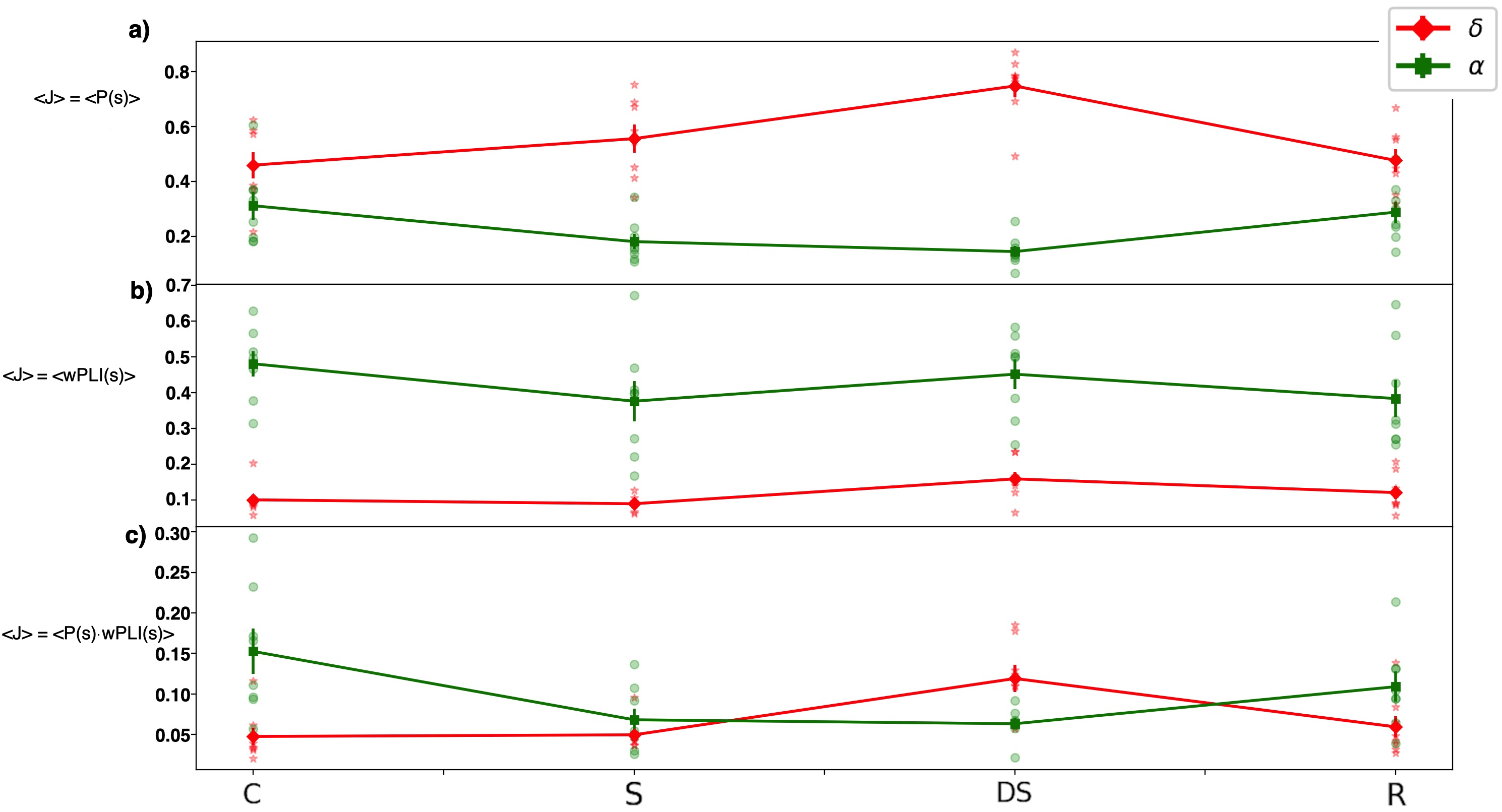}
\caption{\small Average value of the interaction terms, over $4$ different mental states, in Curie-Weiss built using three different approaches. In particular, the interactions have been defined by using \textbf{a}) the power spectrum $\langle J \rangle = \langle P(s) \rangle$, \textbf{b}) the wPLI index, i.e. the correlation $\langle J \rangle =  \langle wPLI(s) \rangle$ between pairs of sensors, and \textbf{c}) the scalar product of the power with wPLI index (i.e. $\langle J \rangle = \langle P(s) \cdot wPLI(s) \rangle$). The legend indicates the considered frequency bands: red line for the $\delta$ and the green line for the $\alpha$, and the related diamonds and squares represent the average value for each label. The error bars have been calculated using the standard error of the mean (SEM). Then, semi-transparent points (i.e. red stars for $\delta$ and green circles for $\alpha$) represent the single individuals. \label{fig:figure_1}}
\end{figure}

The proposed model has been tested with a dataset of EEG signals obtained by recording $8$ healthy volunteers~\cite{main_protocol} undergoing sedation induced by propofol ---see Appendix~\ref{sec:appendix_b} for details.
%
So, each recording started with individuals in the conscious state and terminated after their complete recovery. Four main mental states can be identified: awareness, sedation, deep sedation, and recovery, and for each of them, one sample is extracted from the recording.
The resulting amount of samples (based on $173$ sensors), available for all $5$ frequency bands, allows generating $20$ Curie-Weiss configurations per individual. However, for the reasons reported above, only the $\alpha$ and $\delta$ bands are considered, therefore the number of configurations in this investigation is limited to $8$.
A model configuration has specific values of spin interactions. Let us remind that spins are the mathematical representation of the sensors in the scalp, and their value (i.e. $\pm 1$) is randomly assigned as below described. Instead, the interactions are computed by Eq.~\ref{eq:link}. 
Before proceeding showing results of the simulation, we compare the average value of interactions obtained by using the above equation (i.e. Eq.~\ref{eq:link}) with those (average) values that can be achieved by using the spectral power ($P$) and by the $wPLI$, individually. This comparison is performed over the $4$ mental states and the outcomes have been averaged over all individuals. Results are shown in Figure~\ref{fig:figure_1}, which reports also the related standard error of the mean, i.e. $SEM = \frac{\sigma}{\sqrt{N}}$ with $\sigma$ standard deviation and $N$ number of individuals. A quick inspection of Figure~\ref{fig:figure_1} suggests that our method enhances in good extent the difference between the states $C$ and $DS$, embedding the contribution of the spectral power and that of the wPLI index.
Let us highlight that our goal is now to extract information related to the mental states of individuals by processing the resulting Curie-Weiss models. So, the next step is focused on the identification of the critical temperature for the different configurations of the Curie-Weiss model. This process allows us to build a 'state space' of mental states, where we can quantify, and also visualise, the path followed by individuals across the clinical examination. Then, the paths of mental states are used to train a simple machine learning tool whose goal is to classify conscious states of our individuals.
\subsection{Mental States I: Building the State Space}
\begin{figure}[h!]
\centering
\includegraphics[width=6.0in]{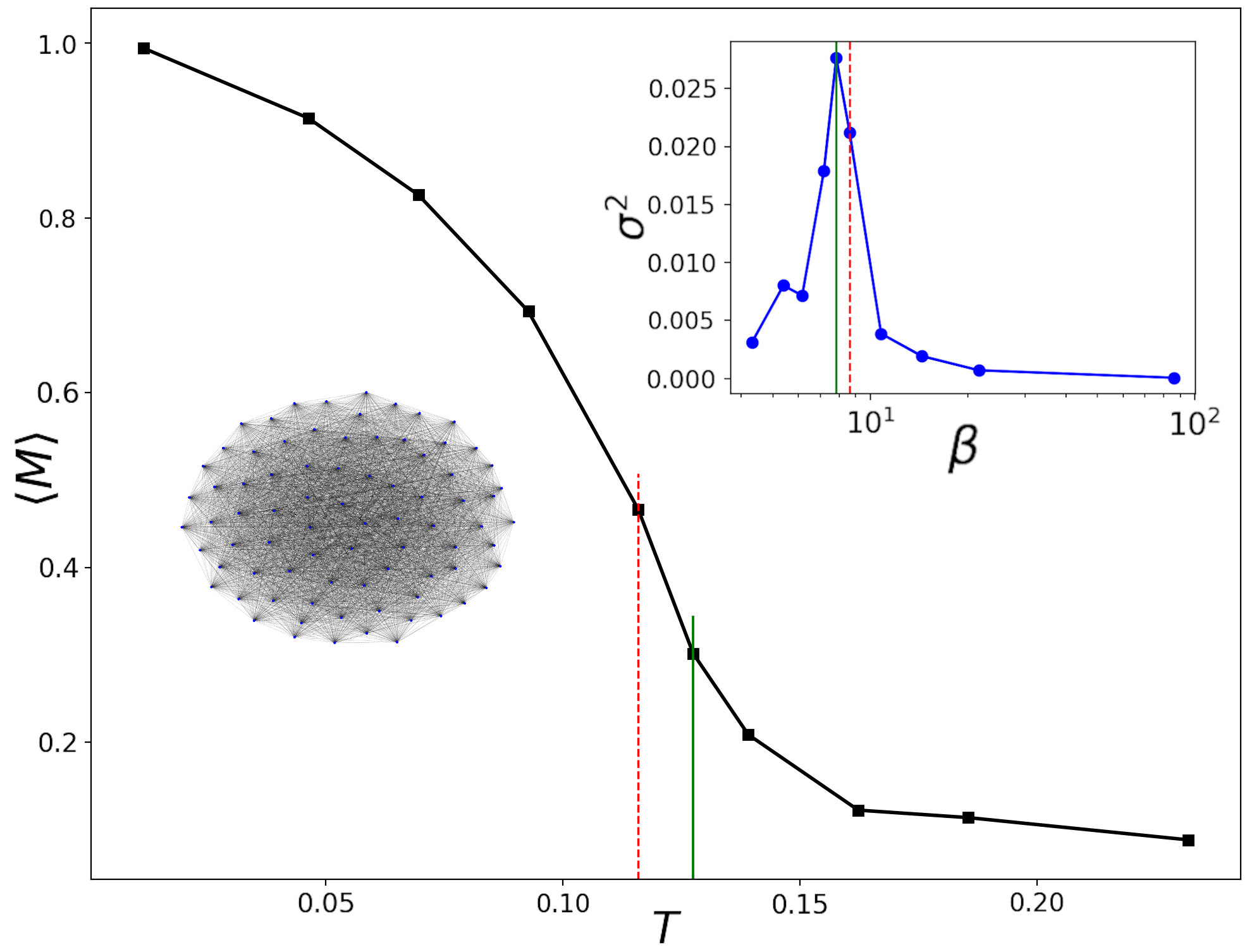}
\caption{\small Average magnetization in function of the system temperature. A pictorial of a Curie-Weiss model is shown close to the main line, while the inset shows the variance of $M$ in function of the inverse temperature $\beta$ (on a semi-logarithmic scale). The critical temperature is indicated by the green line (both in the main picture and in the inset), while the average interaction term for the system ($\langle J \rangle$) is indicated by a red dotted line (both in the main picture and in the inset). \label{fig:figure_2}}
\end{figure}

\begin{figure}[h!]
\centering
\includegraphics[width=6.0in]{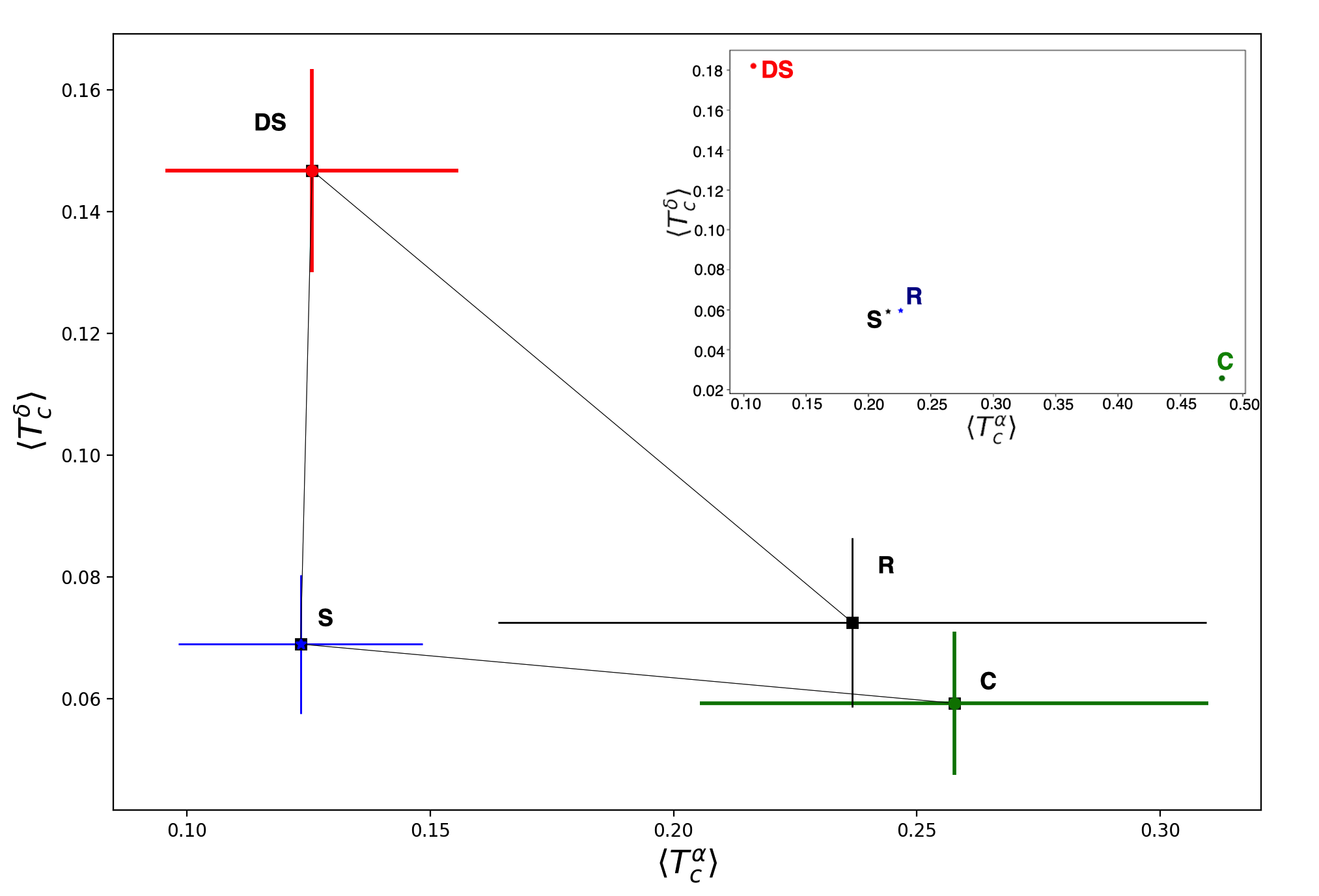}
\caption{\small Diagram $\langle T_c^\alpha \rangle, \langle T_c^\delta \rangle$, with an inset showing the results obtained on a single individual (randomly chosen). Notably, the points in the diagram (and in the inset) represent the $4$ different mental states during the clinical experiment: $C$, $S$, $DS$, and $R$, i.e. consciousness, sedation, deep sedation, and recovery, respectively. The black line indicates the path followed by each individual, undergoing sedation and then recovering to the initial state. \label{fig:figure_3}}
\end{figure}

All numerical simulations have been performed on Curie-Weiss configurations related to every single individual, i.e. spin interactions have not been averaged as (instead) it has been done for the analysis shown in Figure~\ref{fig:figure_1}.
While spin interactions $J$ are \textit{quenched}, we study the dynamics of spins that, at the beginning of each simulation, are randomly set to $\pm 1$. 
Notably, simulations are implemented for studying order-disorder phase transitions, and more specifically for identifying the critical temperature of each configuration. For that purpose, a useful parameter is the absolute value of the average magnetization of the system. 
The latter strongly depends on the system temperature $T$, despite its definition does not include the temperature explicitly, and it reads
\begin{equation}\label{eq:magnetization}
\langle M \rangle = \sum_i^N \frac{|\sigma_i|}{N} 
\end{equation}
\noindent It is worth to observe that beyond identifying the critical temperature $T_c$ for each case, it is possible to assess if $T_c \approx \langle J \rangle$. The latter, as further explained later, can be a relevant feature for designing clinical applications based on the proposed model.
Figure~\ref{fig:figure_2} reports the results obtained on one individual, randomly chosen among those available. To assess whether a temperature is 'critical', the variance of $M$ (indicated as $\sigma^2$) is analysed in function of the inverse temperature $\beta$ since the highest value of $\sigma^2$ is reached at $T = T_c$ ---see inset of Figure~\ref{fig:figure_2}.
Then, once computed the critical temperature for all considered cases, we focus on the numerical difference between these values and the average interaction term (i.e. $\langle J \rangle$) of each Curie-Weiss realisation. In doing so, we found that approximating the $T_c$ with the $\langle J \rangle$ gives small errors, limited to the $12.5 \%$ of $T_c$ (see for instance the lines red and green, indicating the $\langle J \rangle$ and the $T_c$, respectively, in figure~\ref{fig:figure_2}).
Finally, results of the mean-field model are shown in Figure~\ref{fig:figure_3}. Notably, plotting the values of the critical temperatures computed in the band $\delta$ ($T_c^\delta$) versus those computed in the band $\alpha$ ($T_c^\alpha$) allows us to visualise the path from the initial point $C$ to the point $DS$, and that of return. 
So, we have now a state space that contains 'mental paths', whose evolution (or motion) can be further analysed.
Before proceeding to the classification of mental states, we highlight that the order-disorder phase transition studied in the resulting Curie-Weiss model has not a biological meaning in this context. Notably, it is a process simulated to extract information from the resulting model, that we assume can be useful for understanding the dynamics of mental states. In addition, the value of critical temperatures has been computed on a finite size system, while phase transitions occur in the thermodynamic limit. Therefore, further investigations with bigger systems can be useful also to evaluate if the critical temperatures we obtained in our analyses are much different from those one would find scaling the size of the model.
\subsection{Mental States II: Classification}
\begin{figure}[h!]
\centering
\includegraphics[width=5.0in]{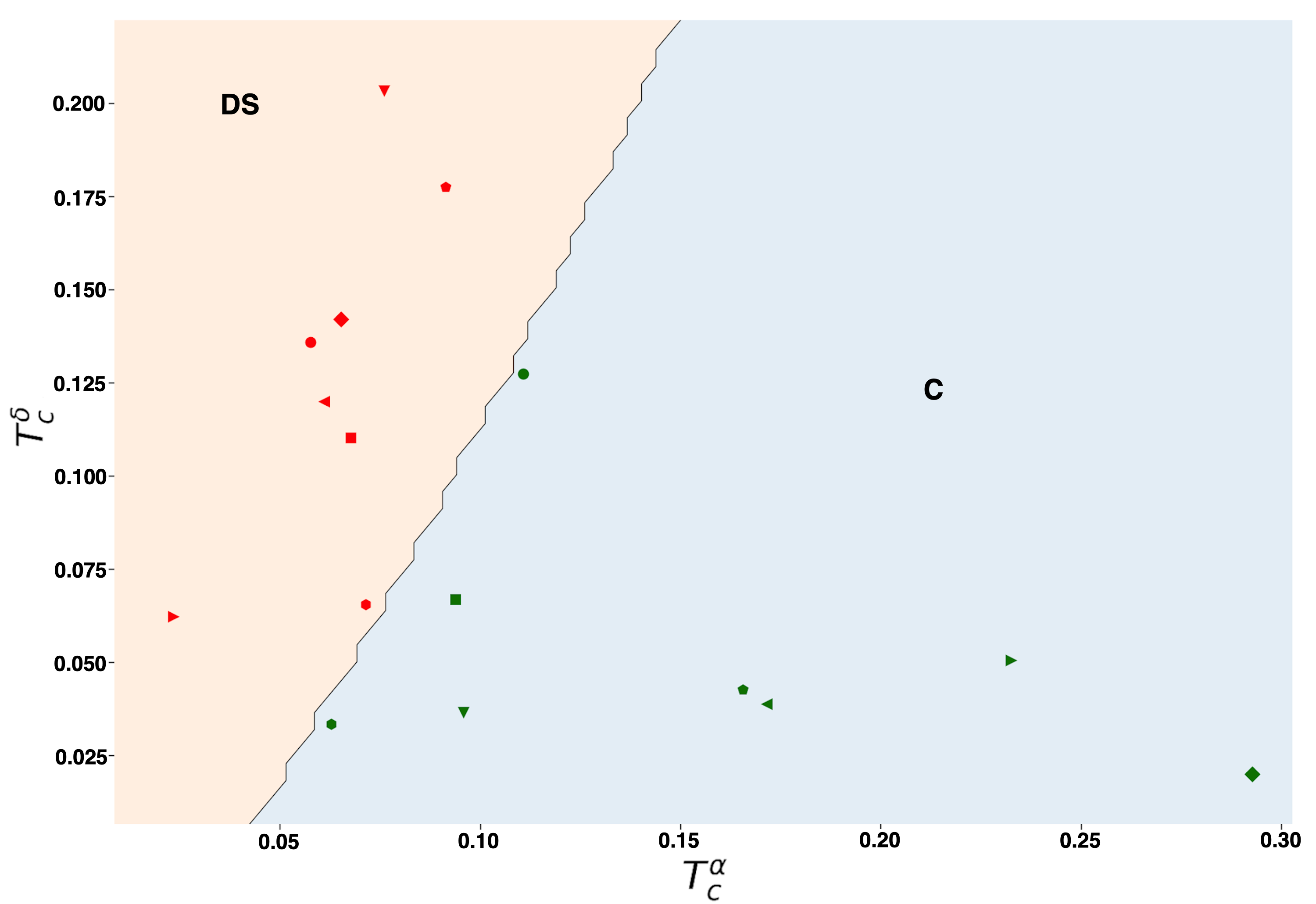}
\caption{\small Results of SVM applied to the EEG dataset on the plane $T_{c}^{\alpha}$ vs $T_{c}^{\delta}$. Different symbols refer to different individuals. The green colour indicates the conscious state ($C$), while the red colour indicates deep sedation ($DS$). The black line identifies the edge between the two classes, i.e. $C$ and $DS$.\label{fig:figure_4}}
\end{figure}

\begin{figure}[h!]
\centering
\includegraphics[width=6.5in]{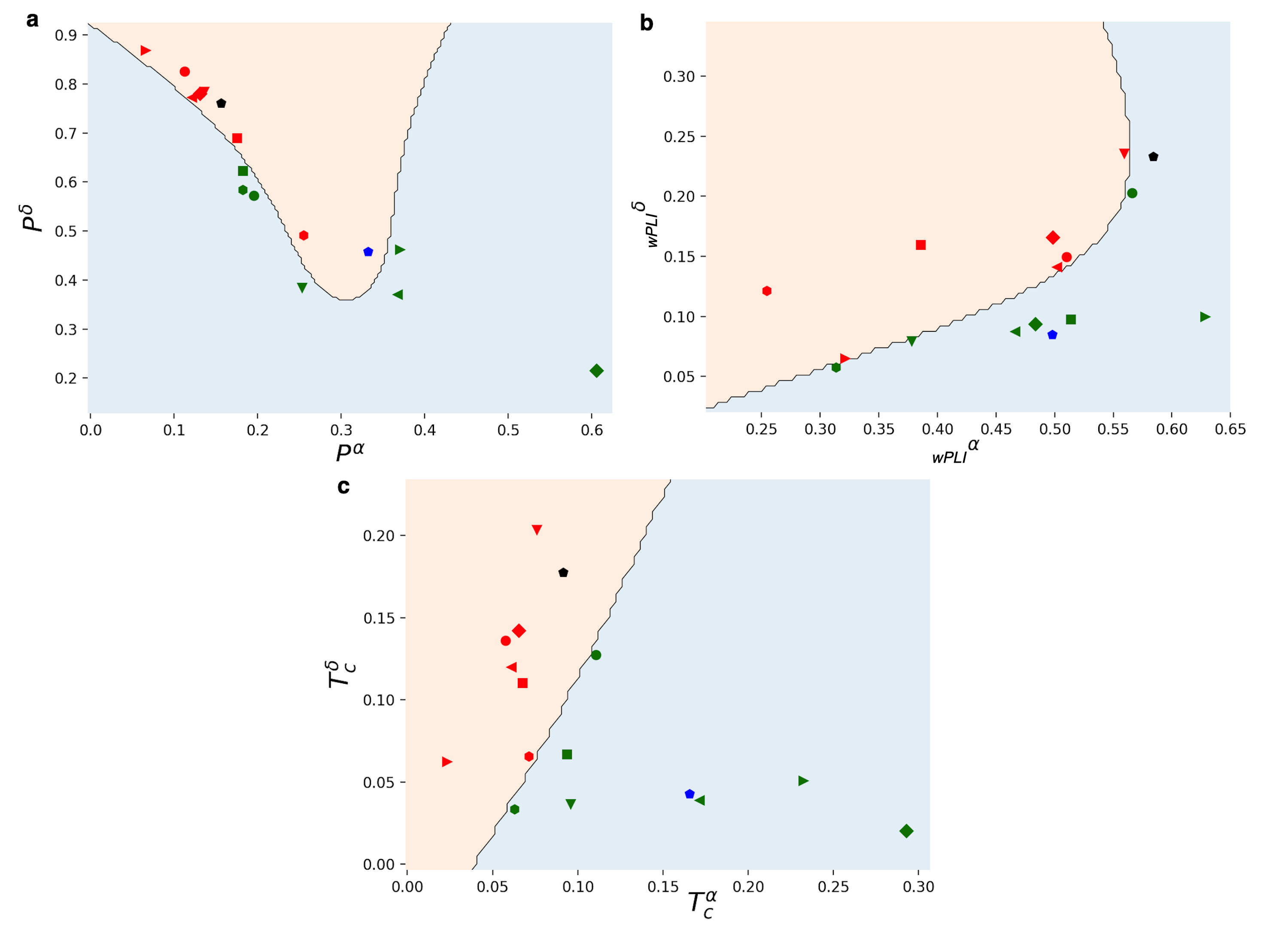}
\caption{\small Comparison between the results of SVM model built with data coming from three different methods: \textbf{a}) Using spectral power; \textbf{b}) Using wPLI; \textbf{c}) Critical temperatures obtained by using Eq.~\ref{eq:link}. Different symbols refer to different individuals. The green colour indicates the conscious state ($C$) and red one that of deep sedation ($DS$). The black line identifies the edge between the two classes, i.e. $C$ and $DS$ in the mental state space. Then, points with the same symbol, but a different colour (e.g. blue and black), indicate an error in the classification process. \label{fig:figure_5}}
\end{figure}

\begin{figure}[h!]
\centering
\includegraphics[width=4.7in]{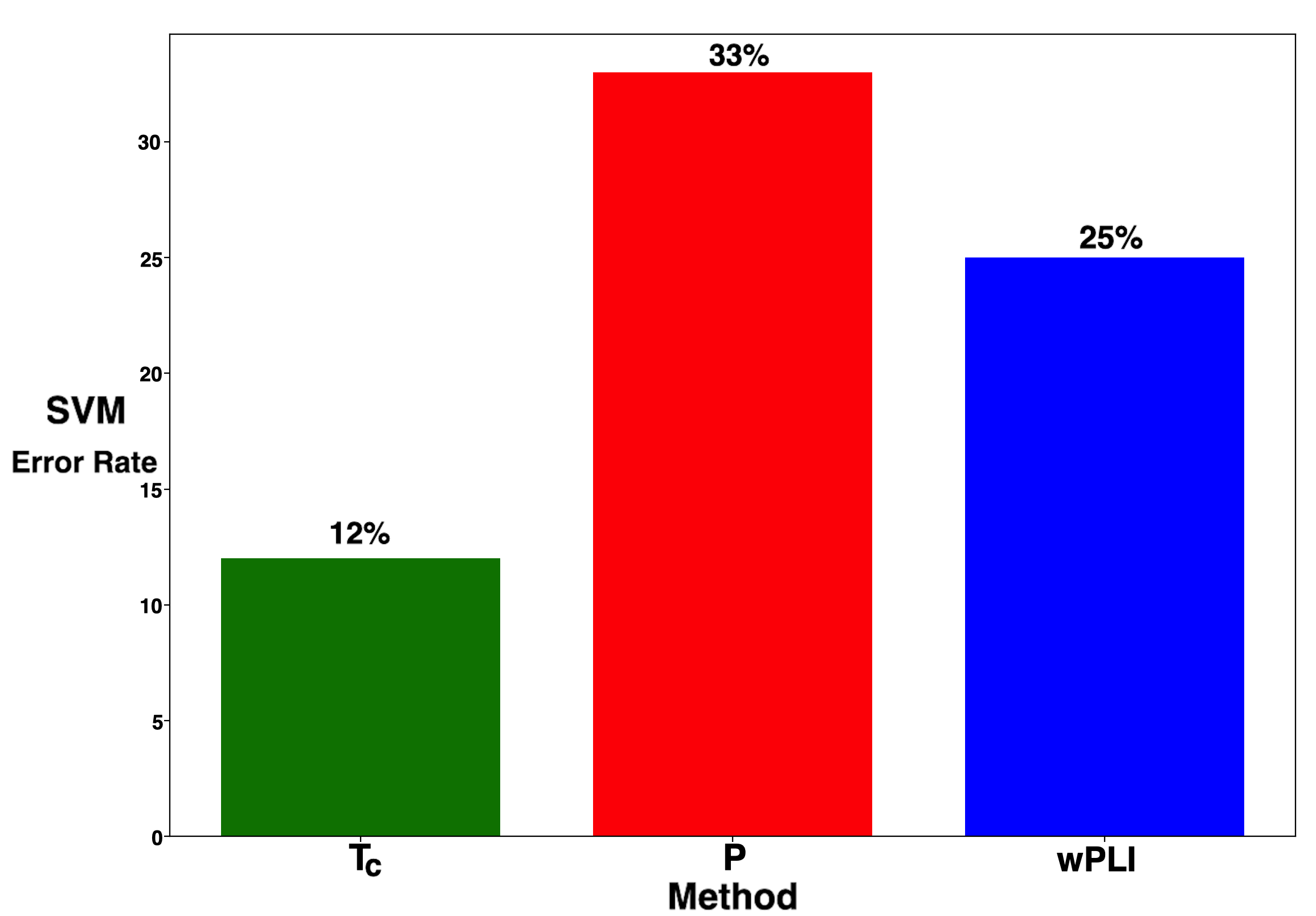}
\caption{\small SVM error rate computed by data coming from three different methods: Critical temperatures ($T_c$) obtained by Eq.~\ref{eq:link}, Spectral power, and wPLI. Note: the lesser the better.  \label{fig:figure_6}}
\end{figure}

Each path obtained by the previous method characterises an individual, thus we use this representation for implementing a Machine Learning tool for classifying mental states.
In our investigation, paths are composed of $4$ points in bi-dimensional state space. Therefore, we can identify a vector, whose entries are the critical temperatures computed for the bands $\alpha$ and $\delta$, for each mental state. The goal is to assess whether these vectors are useful for generating a confident boundary able to separate different mental states. 
So, for the sake of simplicity, we build a classifier for discriminating between the state $C$ and the state $DS$. 
Due to the small size of the dataset, we use an SVM implemented by a kernel based radial basis function (as commonly adopted in classification tasks). 
The outcomes are shown in Figure~\ref{fig:figure_4}, where the axes refer to the two critical temperatures of each individual (i.e. one per frequency band), not to their average values (as in the main plot of Figure~\ref{fig:figure_3}).
Let us briefly describe the procedure for training and testing the SVM. Notably, given a dataset of $8$ individuals, the training has been performed on $7$ out of them, and the testing on the excluded one. This process was then repeated excluding each time one different individual. In doing so, we can evaluate $8$ different testing phase results. 
Following the above procedure, now a further analysis compares the results that an SVM achieves when fed with vectors obtained by three different approaches. Notably, in all cases, vectors resulted from a mean-field model (with the method above described), but its interactions can be defined by Eq.~\ref{eq:link}, or by the power spectrum and the wPLI, individually.
This comparison, shown in Figure~\ref{fig:figure_5}, is quite relevant because we want to evaluate in which extent the utilisation of Eq.~\ref{eq:link} provides an actual benefit for our purposes.
Remarkably, we found that our method produced the smallest error rate in the task of classifying mental states ---see Figure~\ref{fig:figure_6}. In particular, the SVM trained and tested with our method did only one error, over $8$ tests (i.e. $12\%$), while those trained with data coming from the two other methods did more errors (i.e. $33\%$ and $25\%$ by using power spectrum and wPLI, respectively).
Before to conclude this section, we deem important to mention that actual measures might be used to compare outcomes of different models, as the AUC. At the same time in this investigation, considering the number of samples, we found beneficial to evaluate the error rate.

\section{Discussion and Conclusion}\label{sec:conclusions}
In this manuscript, we propose a method for quantifying human consciousness and classifying mental states using EEG signals (see also~\cite{engemann01,srivas03}). Notably, we introduce a mean-field model of the distribution of electrical activity in the brain, whose outcomes are used for training a Machine Learning classifier. 
Inspired by the Tegmark's work~\cite{tegmark01} about the IIT, we study order-disorder phase transitions occurring in Curie-Weiss models whose interactions depend on the phase differences across scalp locations. This analysis allows computing the critical temperatures achieved for different configurations of the model, i.e. on varying the mental state and the considered frequency band of the signal. 
Here, the interactions between spins are computed by the scalar product of the power spectrum with the wPLI index, for each specific band. The benefits coming from this choice are reported in Figure~\ref{fig:figure_5} and Figure~\ref{fig:figure_6}.
Remarkably, the critical temperatures, computed by means of numerical simulations, allow defining a state space where we can observe the path of mental states.
It is worth to mention that, according to Giulio Tononi~\cite{tononi02}, consciousness emerges also during dreamlike phases of the sleep. Therefore, considering previous investigations stating that some individuals reported dreamlike activity, during an induced deep sedation~\cite{gyulahazi01}, both awareness and deep sedation in principle should be considered as conscious states (see also~\cite{yoo01}).
The path between awareness and deep sedation shows, in the middle, the transition states ($S$ and $R$). Thus, summarising, we have two conscious states and two transition states. It is relevant to clarify that we are not using the formal meaning of 'transition state', as usually adopted in stochastic models (e.g. the voter model~\cite{liggett01}), otherwise also the 'deep sedation' state would be defined a transition state since it lasts only for the duration of the drug effect. At the same time, these considerations could be extended further since, for instance, states of coma could be properly classified as absorbing states, and so on. Hence, our definitions have not that level of formality, despite we find interesting to investigate more on this.
So, once defined a mental state space, we use an SVM for tracing the boundary between states $C$ and $DS$.
We remind that the dataset for performing the investigation has been obtained with $8$ individuals wearing EEG sensors. Then, the EEG recordings have been used for building the mean-field model (i.e. Curie-Weiss) and generating a training dataset. The actual choice of the frequency bands, i.e. $\alpha$ and $\delta$, actually depends also on the clinical settings of examinations (e.g. the utilisation of propofol for inducing the sedation).
Results suggest that the critical temperatures are useful to perform classification tasks, discriminating between the two conscious states.
Therefore, thinking about the potential use of the proposed model, at clinical level, we conceive a framework composed of two elements: one devised for computing the average interaction term in different bands, that, as we proved, approximates the critical temperature in the related Curie-Weiss configuration, and the other based on an SVM (or on another Machine Learning algorithm). 
The fact that the average interaction can be approximated by the critical temperature, in principle, is not too surprising. Notably, the critical temperature of the 2D Ising model is $T_c = 1$ for $J=1$. However, it has been worth, for the reasons above described, to confirm that hypothesis.
Beyond the theoretical analysis, which requires further investigations to confirm our achievements, the framework that we conceive could support, after appropriate testing, clinicians in different scenarios. For instance, it could be useful for defining the optimal amount of drug for sedating a patient, or for classifying the level of unconsciousness of unresponsive patients.
Also, it is interesting to observe that the mean-field model realised by EEG recordings might be conceptually related to the 'classical' Tegmark's description of consciousness (i.e. that based on the Ising model). Notably, it would be useful to evaluate how to obtain only one model, across the different bands, and to study its behaviour at different temperatures. Furthermore, we highlight the possibility to extend further this work trying to improve the connection with the IIT framework, in order to develop mathematical tools able to analyse the human consciousness from a perspective supported by the Tononi's insights.
Finally, we deem interesting to provide a comment on the method implemented to represent the path of mental states. Notably, as above reported, each state is described by a vector of critical temperatures (one per frequency band). So, like for other kinds of models, as those based on network theory, the proposed approach can find application in various contexts, in particular when the strength of interaction among the elements of a system is relevant. In our case, the interaction strength clearly shows a time dependency, however that is not a mandatory requirement to implement and analyse a system by means of a Curie-Weiss model. Moreover, the method might provide interesting insights also when built considering interactions defined by meaningful semantic relations among elements of a system, i.e. abstracting from a physical system and therefore increasing its potential applicability to a much wider set of cases.
We conclude emphasising that our results indicate that the EEG signal might be further exploited both for obtaining a deeper understanding of human consciousness, and for implementing novel tools to support clinicians in many complex and critical activities. 
To this aim, there are some important aspects to consider in future investigations. Firstly, here we focused on two frequency bands ($\alpha$ and $\delta$), however it is important to assess if our choice is the optimal one. Notably, although previous literature suggests that these two bands are particularly relevant for the human consciousness, finding the way to exploit also other frequency bands might be useful (as before mentioned, the choice can depend also on the anesthetic drug). Then, interactions among sensors have been identified by means of the wPLI index. However, also other correlation measures could be taken into account. Eventually, we identified transient states, i.e. those states between $C$ and $DS$. We deem that their role, dynamics and properties need to be further analysed, as well as a classification algorithm for their detection could be useful.
\section*{Acknowledgments}
This work was supported by the Belgian National Funds for Scientific Research (FRS-FNRS), the European Union’s Horizon 2020 Framework Programme for Research and Innovation under the Specific Grant Agreement No. 945539 (Human Brain Project SGA3), the University and University Hospital of Liege, the fund Generet, the King Baudouin Foundation, the BIAL Foundation, the AstraZeneca foundation, the Belgian Federal Science Policy Office (BELSPO) in the framework of the PRODEX Programme, the Center-TBI project (FP7-HEALTH- 602150), the Public Utility Foundation ‘Université Européenne du Travail’, “Fondazione Europea di Ricerca Biomedica”, the Mind Science Foundation, and the European Commission. SL is research director at F.R.S-FNRS.

\appendix
\section{Data Classification}\label{sec:appendix_a}
Here we briefly present the algorithm used for classifying mental states, i.e. the Support Vector Machine (SVM). The latter receives as input vectors generated by using the output of the Curie-Weiss model. In doing so, the input vectors contain the critical temperature computed for each frequency band, across the different mental states. For instance, considering the bands $\delta$ and $\alpha$, a vector representing one individual in the conscious state contains the two related critical temperatures, i.e. $V_i(c) = [ Tc^\delta,Tc^\alpha ]$.
The collection of these vectors is used for training and testing the SVM. Notably, since an SVM is a supervised learning model, it requires a training set for the learning process and, usually, such training set is based on a fraction (randomly selected) of the dataset under investigation.
For instance, a generic dataset $D$ contains $n$ elements, i.e. $D = \{(x_1,y_1), (x_2,y_2),...,(x_n,y_n) \}$, so that for each input vector $x_i$ one has the associated output $y_i$, i.e. a label or class of belonging (in our case $x_i$ is a temperature vector, while $y_i$ is the corresponding mental state). 
The dimension of the input vector depends on the number of features which describe the system, so in our case corresponds to the number of considered frequency bands.
Accordingly, a binary classification task can be implemented considering as outputs two possible values, e.g. $+1$ and $-1$. Similarly, for a multi-label case, one has to identify proper values for the output. Then, an SVM aims to identify the 'maximum-margin hyperplane' that separates input vectors according to their corresponding output. In the bidimensional case, the hyperplane is a simple line. The SVM can be both linear and non-linear, depending on the function (defined as Kernel) used to separates vectors into the related classes. Also, the closest vectors to the hyperplane are called 'support vectors'.
For instance, in the linear case, a bidimensional hyperplane can be identified by the equation $\hat{w}\hat{x} + b = 0$, with $\hat{w}$ vector of weights, $b$ (small constant) bias, and $\hat{x}$ feature space of the input vectors. Therefore, the SVM tries to identify the most suitable $\hat{w}$ and $b$ to maximise the margin, i.e. the distance, between vectors belonging to the two (or more) different classes.
Following the above example, in a binary classification, input vectors located above the computed hyperplane belong to one class (e.g. $+1$), while those located below the hyperplane belong to the other class (e.g. $-1$).

\section{Clinical Information}\label{sec:appendix_b}
The full description of the experimental setup implemented to acquire the dataset, used in this investigation, is reported in~\cite{main_protocol}. However, for the sake of completeness, here we provide some relevant details related to the participants and to the experimental protocol.
First of all, we emphasise that the investigation was approved by the Ethics Committee of the Faculty of Medicine of the University of Liege. Also, the group of participants, with mean age $22 \pm 2$ y, includes $4$ males. All participants gave written informed consent, so an appropriate examination, including anamnesis, has been performed to assess potential issues related to the anaesthesia, as pregnancy, mental illness, drug addiction, asthma, and so on. 
Experiments are based on fifteen-minute spontaneous 256-electrode hd-EEG recordings, that can be divided into 4 different states: normal wakefulness, sedation (slower response to command),  loss of consciousness with clinical unconsciousness (i.e. no response to command), and recovery of consciousness.
The total experimental procedure lasted about $5$ hours, including both the positioning of the 256-electrode hd-EEG cap and the recovery time of participants.
\end{document}